\documentclass[preprint,showpacs,showkeys,preprintnumbers,amsmath,amssymb]{revtex4}

\usepackage{graphicx}
\usepackage{amssymb}
\usepackage{dcolumn}
\usepackage[usenames]{color}

\newcommand {\be}{\begin{equation}}
\newcommand {\ee}{\end{equation}}
\newcommand {\bea}{\begin{eqnarray}}
\newcommand {\eea}{\end{eqnarray}}

\newcommand {\EQ}[1]{Eq. (\ref{#1})}

\begin{document}

\title{Diffusive capture processes for information search}

\author{Sungmin Lee}
\author{Soon-Hyung Yook}
\email{syook@khu.ac.kr}
\author{Yup Kim}
\email{ykim@khu.ac.kr}
\affiliation{Department of Physics and
Research Institute for Basic Sciences, Kyung Hee University, Seoul
130-701, Korea}
\date{\today}

\begin{abstract}
We show how effectively the diffusive capture processes (DCP) on
complex networks can be applied to information search in the
networks. Numerical simulations show that our method generates
only $2\%$ of traffic compared with the most popular
flooding-based query-packet-forwarding (FB) algorithm. We find
that the average searching time, $\left<T\right>$, of the our
model is more scalable than another well known $n$-random walker
model and comparable to the FB algorithm both on real Gnutella
network and scale-free networks with $\gamma =2.4$. We also
discuss the possible relationship between $\left<T\right>$ and
$\left<k^2\right>$, the second moment of the degree distribution
of the networks.
{} %
\end{abstract}
\maketitle


The study on the complex networks has been attracted many
researchers in diverse fields \cite{3ves_book,11BA}. Due to the
inherent complexity in their structures, the dynamical properties
of many physical systems on the networks have shown rich behaviors
which depend on the structure of the underlying networks and are
far from the mean-field expectations \cite{24mendes_critical}. The
diffusive capture process (DCP) on complex network is one such
example \cite{12our_cma,13our_ll}.
The DCP or the moving trap model has been studied in many physics
literatures \cite{15redner_dcp} on chemical kinetics, wetting,
melting, etc. This process can be mapped into coupled random walks
(RWs), in which $n$ predator random walkers (lions or moving trap)
stalk a prey random walker (a lamb or a moving particle). On a
$d$-dimensional regular lattice, the survival probability $S(t)$
of the lamb is given by
\cite{21feller_book,22redner_book,23weiss_book}
\begin{equation}
S(t) \sim \left\{
\begin{array}{lll}
t^{-\beta} & for & d < 2\\
(\ln{t})^{-n} & for & d = 2\\
\mbox{finite} & for & d > 2,
\end{array}\right.
\end{equation}
where the exponent $\beta$ varies as $n$ and the ratio of
diffusion constant of walks. Like many physical systems on complex
networks, we have found that the $S(t)$ of the DCP on complex
networks also deviates from the mean-field expectation depending
on the degree distribution of the underlying networks. This
anomalous behavior has been found to come from the existence of
dominant hubs in complex networks \cite{12our_cma,13our_ll} and
can be applied to information search in complex networks.

The searching information in complex networks, such as
Peer-to-Peer (P2P) networks and the world-wide-web (www), using
random walkers \cite{Kleinberg,7huberman_search} is an interesting
and important application in diverse areas including physics,
computer science, and neuroscience
\cite{7huberman_search,equiluz,Braitenberg}. The efficiency of the
information search can be defined by two factors, 1) amount of
traffic at time $t$ defined by the total number of packets
existing on the network at $t$ and 2) searching time which
corresponds to the average life time of a lamb in the DCP. In
general, these two factors are competing to each other, i.e.,
reducing the traffic congestion causes long waiting time to find a
given information. Therefore, it is very important and difficult
to design a less traffic congesting model with short searching
time. In this paper, we consider the \textit{pure} P2P networks as
an example of information search by taking notice of the fact that
the most of traffic generated by P2P applications consists of the
query packets to find out the node which has the requested file
out of the P2P jungle \cite{3ves_book}. A \textit{pure} P2P
network does not have the notion of clients or servers, but only
equal peer nodes that simultaneously function as both clients and
servers to the other nodes on the network. Thus, there is no
central server managing the network and no central router. By
applying the anomalous behavior of the DCP on complex networks,
%
we drastically reduce the traffic congestion. We
will also show that our model can provide implementable searching
time.
We expect that our results can be easily generalized to apply many
other information search problems and communication networks, such
as neural networks.

The performance of searching algorithm is crucially affected by
the underlying topology \cite{4up2p}. Many studies on the
large-scale topology of P2P networks have uncovered that the
probability distribution of a node with degree $k$ follows the
power law, \bea \label{pk} P(k) \sim k^{-\gamma}, \eea with
$\gamma < 3$ \cite{3ves_book,4up2p,7huberman_search,9mapping_gnu},
or highly skewed fat-tailed distributions \cite{5snapshot_gnu}.
The network satisfying \EQ{pk} is called a scale-free (SF) network
\cite{11BA}. In these networks with $\gamma<3$, several nodes have
most of degrees or connections. These nodes are called hubs and
many important properties of complex networks are dominated by
them \cite{3ves_book,11BA}. However, most of the popular
algorithms used in many P2P networks do not take advantage of the
underlying structure. In those algorithms a query packet is
generated by a node and forwarded to the nearest neighbors until
it finds the requested information during a query event. For
example, the flooding-based query-packet-forwarding (FB) algorithm
\cite{4up2p}, which is used in BitTorrent
\cite{overlay,bittorrent}, spreads the query packets to all nodes
within a pre-assigned diameter. Thus, this algorithm causes
significant traffic congestion. The $n$-random walker ($n$-RW)
model, which is used in LMS (Edutella) \cite{overlay,edutella}, is
another well studied model for searching information
\cite{4up2p,7huberman_search}. $n$-RW model can cause long waiting
time because of the dynamical properties of RWs on complex
networks \cite{14noh}. Gnutella and Kazaa use both algorithms
\cite{gnutella,kazaa}. Other P2P applications are structured and
use global information such as distributed hash table (DHT)
\cite{overlay,dht}. Thus, the traffic generated by those P2P
applications using global information can be ignorably small
compared with the \text{pure} P2P applications.
%
However, pure P2P protocols mentioned above and
their clones are still used very widely and listed in the most
popular P2P protocols (for example see the Ref.
\cite{mostpop_p2p}.)
%
Therefore, we focus on the searching algorithm of \textit{pure}
P2P networks without global information in this paper. Inspired by
our recent discoveries in the DCP \cite{12our_cma,13our_ll}, we
introduce a new model for information search in which not only the
query packets but also the information packets take random walks
on the network. We will show that our new model has two main
benefits compared to the other algorithms: 1) the amount of
traffic is always constant and much less than FB algorithm. 2)
Much less searching time than that for $n$-RW model. Using these
two benefits, we expect that our algorithm can provide more
optimized algorithm compared to FB or $n$-RW for \textit{pure} P2P
networks.

We now explain our model based on the anomalous behavior of DCP on
the complex networks in detail. In the model, each node sends out
an information packet whose main part, for example, consists of
names of files stored in it, without regard to the existence of
query events. Each of these packets takes random walks along the
network connections. Thus, if there are $N$ nodes in the network
then $N$ information packets take random walks. Independently, a
randomly chosen node sends out one query packet to find a specific
file. The query packet also takes random walks. The special
feature in our model is thus in the fact that not only the
querying packet moves but the information of all files moves on
the network. If the query packet meets an information packet which
has the requested file name in its list, then the random walk of
the query packet is terminated but the information packet
continues random walks for the next query. Thus, our model for
information search can be mapped into the DCP, i.e., the query
packet and the $N$ information packets correspond to the one
random walking lamb and $N$ random walking lions, respectively.
Therefore, we will call our model the $N$ lions and one lamb (NLL)
model.

We assume $n_f$ available files on the given network. For the
distribution of files and frequency of queries, we use two kinds
of file distribution and frequency of queries. In the first kind,
we assume that each node of the network has one randomly chosen
file among $n_f$, and sends out an information packet with the
name of the file stored in it, its IP address, etc. And a randomly
chosen node sends out a query packet to find one randomly chosen
item among $n_f$ files. Thus the popularity of files and frequency
of queries are uniform. In the second kind which is invoked by
more realistic P2P network \cite{zipfbook}, the popularity of
files or the probability  to find the $r$-th most popular file in
the network is assumed to be proportional to $1/r$ (Zipf's law)
\cite{QLv1,zipf,QLv2}. The frequency of queries to find the $r$-th
most popular file is also assumed to be proportional to $1/r$
which is consistent with some empirical observations
\cite{QLv1,zipf,QLv2}. We call the first kind uniform distribution
and the second kind Zipf-distribution from now on. The empirically
obtained Zipf-distribution also implies that it is more probable
to increase the number of more popular files rather than new or
rare files when a new user joins the P2P network. Thus, we assume
that $n_f$ is fixed for each given network for simplicity.

In the simulation we use two kinds of networks. One is the
theoretical SF network with $\gamma=2.4$ to mimic the virtual P2P
networks. The SF networks are generated by the method suggested by
Goh et al. \cite{28goh}. To compare the scalability of each model
or algorithm, we control the number $N$ of nodes in the networks
as $N=10^3\sim 10^6$. The other kind is the snapshot of a real
Gnutella topology obtained from Refs. \cite{5snapshot_gnu}, which
has $1,074,843$ ($\approx 10^6$) nodes. For comparison with the
results on the theoretical SF networks, we extract the
sub-networks of Gnutella from the huge snapshot without changing
the topological properties. Constructing sub-network with $N<10^6$
from the given snapshot of the Gnutella network without changing
the topological properties is not trivial one
\cite{29jeong_samplednet,sample_rs,sample_ours}. To extract
Gnutella sub-networks having $N=10^3 \sim 10^5$ nodes from the
snapshot, we use a RW; place a particle at a randomly chosen node
and let the particle take random walks until it visits $N$
different nodes. Then construct sub-networks with these $N$
visited nodes and the links which connect any pair of nodes among
the $N$ visited nodes in the original network. We have verified
that the resulting sub-networks and original Gnutella network have
almost the same degree distribution, degree-degree correlations
and hierarchical structures \cite{sample_ours}. All quantities
measured in the simulations are averaged over $10$ network
realizations and $100$ different histories for each network
realization.

%
\begin{figure}
\includegraphics[width=8.5cm]{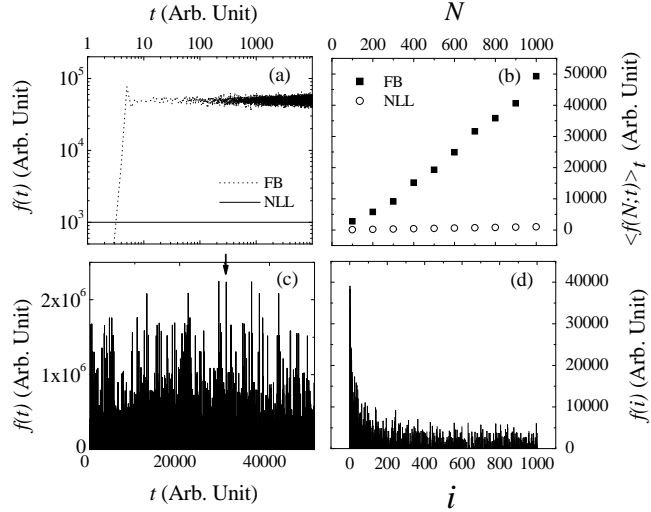}
\caption{(a) Plot of traffic $f(t)$ against $t$. The average
traffic of NLL model is calculated from simple theoretical
arguments. (b) Time averaged traffic $\left<f(N;t)\right>_t$
generated by FB and by NLL  for various network sizes $N=10^2 \sim
10^3$. (c) The time evolution of $f(t)$ obtained from a single run
of simulation of FB algorithm. One of the local maxima of traffic
generated by FB algorithm is marked by the arrow. (d) The traffic
of FB algorithm measured at each node $f(i)$ when $f(t)$ has the
local maximum.}
\end{figure}
In Fig. 1(a), we compare the traffic $f(t)$ generated by FB
algorithms to that by NLL model on the SF networks with $N=10^3$
nodes. At each time $t$, all packets whose pre-assigned
time-to-live (\textit{TTL}) counter is larger than 0 are forwarded
to the nearest neighbors. The \textit{TTL} of each packet
decreases by one when the packet is forwarded to its nearest
neighbors, and if $TTL=0$ then the query event is forced to end
\cite{3ves_book}. The uniform distribution for files and frequency
of queries is used. We assign $n_f=5$ and \textit{TTL}$=6$ for FB
algorithm. The different values of the $n_f$ and $TTL$ give us the
similar results. In order to prevent the overflow caused by a
large number of packets in FB algorithm, we assume that a new
query event can occur when one of the query packets succeed to
find the requested file. In this case, the other query packets,
which fail to find the requested file, are forwarded until their
\textit{TTL}s become $0$. If all the query packets fail to find
the requested file, then a new query event can occur only when the
\textit{TTL}s of the previous packets are expired. The traffic
generated by FB algorithm during each query event is known to
increase exponentially as
\begin{equation}
f(t=TTL)\approx \left< k \right> \left(
\frac{\left<k^2\right>-\left<k\right>}{\left<k\right>}
\right)^{TTL-1} ,
\end{equation}
where $\left<k\right>$ and $\left<k^2\right>$ are the first and
second moments of network degree distribution, respectively
\cite{3ves_book}. However, the traffic of NLL model is always
$N+1(=1001)$ for successive query events. The simulation result
shows that FB algorithm generates around $50$ times more traffic
than NLL model on the average. If there are $q$ simultaneous query
events, then the average traffic for FB algorithm increases simply
by $q$ times of the average traffic shown in Fig. 1(a), but it
becomes simply $q+N$ for NLL model. The successive single query
events of $n$-RW model produce the smallest traffic among the
algorithms considered in this paper. Since the traffic of $n$-RW
model is $qn$ for $q$ simultaneous query events, if $q=N$, i.e.,
every node in the network sends out a query packet simultaneously,
then the traffic generated by $n$-RW model can exceed the traffic
of NLL model depending on the value of $n$. We also consider the
dependence of $f(t)$ on the network size $N$, $f (N;t)$. Fig. 1(b)
shows the time averaged traffic generated by FB algorithms,
$\left< f_{FB} (N;t) \right>_t$, and by NLL model, $\left< f_{NLL}
(N;t) \right>_t$, on the SF networks with $N=10^2 \sim 10^3$. As
$N$ increases, $\left< f_{FB} (N;t) \right>_t - \left< f_{NLL}
(N;t) \right>_t$ considerably increases. Fig. 1(c) displays the
time evolution of the traffic obtained from a single run of
simulation for FB algorithm. It implies a practical importance.
Due to the large fluctuations, for example, around at $t \approx
3.0 \times 10^4$ (marked by the arrow) the traffic on the network
can have local maxima exceeding $2 \times 10^6$, which is $2,000$
times larger than the traffic generated by NLL model. At the
moment of occurring such large amount of traffic, FB algorithm can
causes severe traffic congestion over the network. However, NLL
model always guarantees a constant level of traffic, which is much
less than that of FB algorithm and comparable to that of $n$-RW
model.

Since the probability that a RW visits a node of degree $k$ is
given by $p_v (k) = \frac{k}{\sum^N_{i=1} k_i}$
\cite{21feller_book,14noh}, the average traffic of NLL model at
the node having $k$ links can be estimated by $\left<f (k)\right>
= N\frac{k}{\sum^N_{i=1} k_i}$. Therefore, the hub can have
considerable amount of lion-traffic in the NLL model. In order to
find the bottleneck in FB algorithm, we measure the traffic of
each node $f(i)$ when $f(t)$ has the local maximum for FB
algorithm (Fig. 1(d)). By definition of the static model
\cite{28goh}, the smaller node index $i$ has the larger $k$. From
the data, we find that the traffic of the largest hub ($i=1$)
reaches around $4\times 10^4$ which is much larger than the
average traffic of NLL model at the largest hub $\left<
f(k_{max})\right>$ or the possible maximum traffic of NLL model
($f_{max} = 1001$ for $N=1000$). The maximum traffic of a link at
time $t$ can be estimated by mean-field type arguments as $f
(k_{max}; t) + \left<f (t) \right>$ for FB algorithm which is much
larger than that for $n$-RW and NLL model, $\frac{f (k_{max};
t)}{k_{max}} + \frac{\left<f (t) \right>}{\left<k\right>}$. Here,
$f (k_{max}; t)$ represents the traffic on the largest hub at $t$.

%
\begin{figure}
\includegraphics[width=8.5cm]{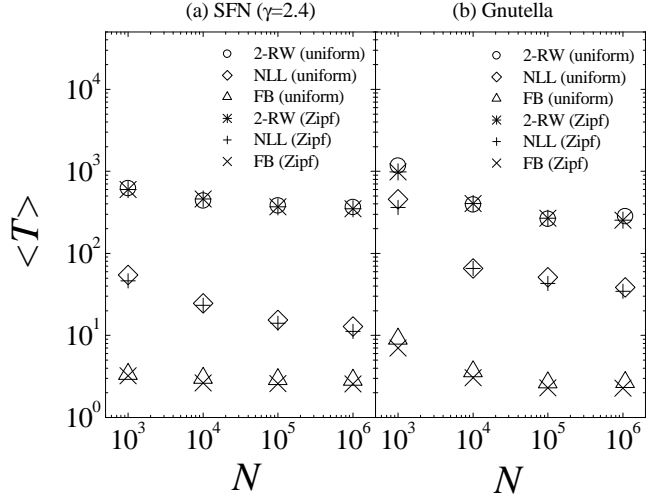}
\caption{Log-log plot of the average time $\left<T\right>$ taken
to find the requested information with a fixed value of
$n_f$($=500$) on SF networks with $\gamma=2.4$ (a) and real
Gnutella networks (b). $\left<T\right>$ decreases as increasing
$N$ both for $2$-RW and NLL models, while $\left<T\right>$ for FB
algorithm remains almost constant on SF networks (a) and decreases
on real Gnutella networks. Open symbols: both the frequency of
queries and popularity of files follow the uniform distribution.
Crossed symbols: both the frequency of queries and popularity of
files follow the Zipf-distribution.}
\end{figure}
In Fig. 2 we show the average searching time, $\left< T \right>$,
for each algorithm. The searching time $T$ is defined by the time
taken to find the requested information, and thus, it corresponds
to the life time of a lamb in DCP. In the simulations we use
infinite \textit{TTL}s. Since the searching time of $n$-RW model
depends on the value of $n$, we need a criterion for $n$. Here we
use the condition $f_{n-RW} (t) = f_{NLL} (t)$ when $q=N$, so we
fix $n=2$ in the following simulations. In Fig. 2, we display the
average searching time when $n_f =500$. From the data we can not
find any significant difference between uniform distribution and
Zipf distribution of files and queries. The average searching time
of NLL model on SF networks is, at least, $10$ times faster than
$n$-RW model on SF networks. For example,
$\left<T\right>_{2-RW}/\left<T\right>_{NLL}=365.417/12.84 \approx
28$ on SF networks for $N=10^6$. However,
$\left<T\right>_{NLL}/\left<T\right>_{FB}=12.84/2.9 \approx 4.4$
on the same size of SF networks (see Fig. 2(a)). Therefore, the
average searching times satisfy the inequality,\bea
\label{inequal} \left<T\right>_{FB} < \left<T\right>_{NLL} <
\left<T\right>_{2-RW} \eea for all $N \leq 10^6$.

Note that $\left<T\right>$ of NLL model decreases much faster than
that of $n$-RW model as increasing $N$ and approaches to
$\left<T\right>_{FB}$, i.e., the difference between NLL and $2$-RW
models, $\left<T\right>_{2-RW}-\left<T\right>_{NLL}$, increases
while $\left<T\right>_{NLL}-\left<T\right>_{FB}$ decreases as
increasing $N$ (see Fig. 2(a)). We find the same behavior of
$\left<T\right>$'s on the real Gnutella (sub-) networks (see Fig.
2(b)). This can be understood from the dynamical properties of RWs
on complex networks. Since the probability that a RW visits a node
with the degree $k$ is proportional to $k$
\cite{21feller_book,14noh}, the probability that a query packet
finds a requested file at a node with degree $k$ is proportional
to $k$ in $n$-RW model. But in NLL model, due to the random
walking information packets, the probability is proportional to
$k^2$\cite{13our_ll}. As a result, the hubs in the network can
collect more packets in our NLL model than $n$-RW model, and thus
become effective attractors. This effect becomes enhanced if the
second moment of degree distribution $\left<k^2\right>$ diverges,
or for SF networks with $\gamma <3$. In SF networks the degree of
the hub increases with $N$ \cite{3ves_book}. Therefore, both
$\left<T\right>_{NLL}$ and $\left<T\right>_{2-RW}$ decrease but
$\left<T\right>_{2-RW} - \left<T\right>_{NLL}$ increases as
increasing $N$.
These provides an indirect evidence that the $\left< k^2 \right>$
grows more rapidly than $\left<k\right>$ as increasing $N$ in the
Gnutella network.

In this paper we introduce a new model for information search,
which can be easily applied to many information search problems
such as P2P file sharing networks.
%
By numerical simulations, we verify two important
benefits of our model: 1) first of all, our model can drastically
decreases the traffic congestion compared to FB algorithm and 2)
it can provide more implementable scalability in average searching
time than $n$-RW model.
Because of these two benefits, NLL algorithm suggested
in this paper can be utilized to optimize information search on
\textit{pure} P2P networks. The dynamical properties of NLL model
can be easily understood from the DCP and the dynamical properties
of RWs on complex networks. Since the probability that a RW visits
a node with degree $k$ is proportional to $k$
\cite{21feller_book,14noh}, $P(k)$ becomes relevant. If $\left<
k^2 \right>$ diverges then most of the information would be
gathered at several hubs without the knowledge of the global
information on the distribution of files. Therefore, the hubs
spontaneously play a very similar role of the directory servers in
structured P2P networks \cite{4up2p} which provide better
scalability than $n$-RW model.

Though P2P network is virtual network, nodes with large number of
connections in real network can have large number of degree in P2P
network. In certain real P2P networks such as Gnutella, the nodes
which satisfy some requirements such as sufficient network
resources, CPU speed, etc. are prepared to be the hubs with a
large number of degree \cite{gnutella}. For simplicity, we assume
the degree of virtual P2P network is approximately proportional to
the degree of real network and each node and link can treat an
unlimited amount of packets at each time step. However, the
detailed studies with more realistic restrictions such as finite
buffers, different data transfer rate and \textit{ad hoc}
properties of network topology are necessary
%
for further studies to apply NLL model to the real P2P
applications.

Finally, some information transfer networks, such as neural
networks, satisfy \EQ{pk} with $\gamma<3$ \cite{equiluz} or have
huge hub like the Gnutella network \cite{5snapshot_gnu}. In these
networks, we expect that the networks have self organized their
structure to improve the efficiency for finding information such
as reducing the traffic or the searching time. By combining our
results, we expect that the non-mean-field type behavior of DCP
can give a clue to the emergence of SF structure or formation of
hubs in information transfer networks observed in nature
\cite{equiluz, Guimera}.

We thank to Dr. Kwon and Dr. Yoon for useful discussions. This
work is supported by Korea Research Foundation Grant No.
KRF-2004-015-C00185.


\end{document}